\documentclass{IOS-Book-Article}
\def\Z{\mathbb Z}
\def\R{\mathbb R}
\def\N{\mathbb N}
\def\A{\mathcal A}

\usepackage{times}
\normalfont
\usepackage[T1]{fontenc}
\usepackage{amssymb,amsthm,amsmath,graphics}
\newtheorem{thm}{Theorem}
\newtheorem{prop}[thm]{Proposition}

\theoremstyle{definition}
\newtheorem{de}[thm]{Definition}
\newtheorem{ex}{Example}
\newtheorem{claim}{Claim}

\begin{document}
\begin{frontmatter}                           

\title{Quasicrystals: algebraic, combinatorial \\and geometrical aspects%
}
\runningtitle{Quasicrystals}

\author[A]{\fnms{Edita} \snm{Pelantov\'a}%
\thanks{Correspondence to: Edita Pelantov\'a, Department of Mathematics, Faculty of Nuclear Sciences
and Physical Engineering, Czech Technical University, Trojanova
13, 120 00 Praha 2, Czech Republic. Tel.: +420 224 358 544; Fax:
+420 224 918 643; E-mail: pelantova@km1.fjfi.cvut.cz.}},
\author[A]{\fnms{Zuzana} \snm{Mas\'akov\'a}}
\runningauthor{E. Pelantov\'a et al.}
\address[A]{Doppler Institute for Applied Mathematics and Mathematical
Physics\\and\\ Department of Mathematics, FNSPE, Czech Technical
University}

\begin{abstract}
The paper presents mathematical models of quasicrystals with
particular attention given to cut-and-project sets. We summarize
the properties of higher-dimensional quasicrystal models and then
focus on the one-dimensional ones. For the description of their
properties we use the methods of combinatorics on words.
 \end{abstract}

\begin{keyword}
Quasicrystals\sep cut-and-project set\sep infinite word\sep
palindrome\sep substitution
\end{keyword}
\end{frontmatter}

\section{Introduction}

Crystals have been admired by people since long ago. Their
geometrical shape distinguished them from other solids. Rigorous
study of crystalline structures started in years 1830--1850 and
was crowned around 1890 by famous list of Fedorov and Schoenflies
which contained 32 classes of crystals. Their classification was
based purely on geometry and algebra. The simplest arrangement,
arrangement found in natural crystals, is a simple repetition of a
single motive. In mathematics, it is described by the lattice
theory, in physics, the subject is studied by crystallography.
Repetition of a single motive means periodicity. Another
remarkable property, characteristic for crystals, is their
rotational symmetry, i.e.\ invariance under orthogonal linear
transformations.

Important consequence of the lattice theory is that neither planar
nor space (three-dimensional) periodic structures can reveal
rotational symmetry of order 5 or higher than 6,
see~\cite{Senechal}.

The discovery made by  Max von Laue in 1912 enabled the study of
the atomic structure of crystals via X-ray diffraction patterns
and, in fact, justified the theoretical work developed by
crystallography before. In case of periodic structures, the type
of rotational symmetry of the crystal corresponds to the type of
rotational symmetry of the diffraction diagram.

The discovery that rapidly solidified aluminium-manganese alloys
has a three-dimensional icosahedral symmetry, made by Shechtmann
et al.~\cite{Shechtman} in 1982, was therefore an enormous
surprise for the crystallographic society. The diffraction diagram
of this alloy revealed five-fold rotational symmetry. Materials
with this and other crystallographically forbidden symmetries were
later fabricated also by other laboratories by different
technologies. They started to be called quasicrystals.

Schechtman's  discovery shows that periodicity is not synonymous
with long-range order. The definition of long-range order is
however not clear. By this term crystallographers usually
understand ordering of atoms in the material necessary to produce
a diffraction pattern with sharp bright spots. This is also used
in the general definition of crystal adopted by Crystallographic
Union at its meeting in 1992.

The only clear requirement agreed upon by all scientists is that
the set modeling quasicrystal, i.e.\ positions of atoms in the
material, should be a Delone set. Roughly speaking, this property
says that atoms in the quasicrystal should be `uniformly'
distributed in the space occupied by the material. Formally, the
set $\Sigma \subset \mathbb{R}^d$ is called {\it Delone} if
 \begin{description}
 \item[(i)] (uniform discreteness):  there exists $r_1>0$ such that
 each ball of radius $r_1$ contains at most one element of
 $\Sigma$;
 \item[(ii)] (relative density):  there exists $r_2>0$ such that
  each ball of radius $r_2$ contains at least one element of
  $\Sigma$.
\end{description}

The requirement of the Delone property is however not sufficient,
for, also positions of atoms in an amorphous matter form (a
section of) a Delone set. Therefore Delone sets modeling
quasicrystals must satisfy other properties. According to the type
of these additional requirements there exist several approaches to
quasicrystal definitions~\cite{lagarias-finite,Lagarias}: The
concept of Bohr (Besicovich) almost periodic set, is based on
Fourier analysis. The second concept of Patterson sets is based on
a mathematical analogue of X-ray diffraction, developed by Hof.
The third concept, developed by Yves Meyer, is based on
restriction on the set $\Sigma - \Sigma$ of interatomic distances.
It is elegant and of purely geometric nature: A {\it Meyer} set
$\Sigma \in \mathbb{R}^d$ is a Delone set having the property that
there exists a finite set $F$ such that
$$
\Sigma - \Sigma \subset \Sigma +F\,.
$$
In \cite{lagarias-finite}, Lagarias has proven that a Meyer set
can equivalently be defined as a Delone set $\Sigma$ such that
$\Sigma-\Sigma$ is also Delone.

There exists a general family of sets $\Sigma$ that are known to
have quasicrystalline properties: the so-called cut and project
sets, here abreviated to $C\&P$ sets. Various subclasses of these
sets appear to satisfy all three above quoted definitions of
quasicrystals.

The paper is organized as follows: The construction of
quasicrystal models by cut and projection is introduced in
Section~\ref{sec:cap} and illustrated on an example of
cut-and-project set with 5-fold symmetry in Section~\ref{sec:rot}.
The remaining part of the paper focuses on the properties of
one-dimensional cut-and-project sets. Section~\ref{sec:def}
provides their definition Sections~\ref{sec:eqv},~\ref{sec:compl}
and~\ref{sec:subst} show their diverse properties, both geometric
and combinatorial.

\section{Cut-and-project Sets}\label{sec:cap}

The construction of a cut-and-project set ($C\& P$ sets) starts
with a choice of a full rank lattice: let $x_1,x_2,\ldots, x_d\in
\mathbb{R}^d$ be linearly independent vectors over $\mathbb{R}$,
the set
$$
L= \left\{a_1x_1 + a_2x_2+ \ldots + a_dx_d \ \mid \ a_1, \ldots,
a_d \in \mathbb{Z}\right\}\,.
$$
is called a {\it lattice}. It is obvious that a lattice is a
Delone set. Mathematical model for ideal crystal (or perfect
crystal) in $\mathbb{R}^d$ is a set
   $\Lambda$, which is formed by a finite number of shifted copies
   of a single lattice $L$. Formally, $\Lambda$ is a perfect crystal if
$\Lambda = L+ S$, where $S$ is a finite set of translations.

Since lattices satisfy $L -L \subset L$  and a perfect crystal
satisfies $\Lambda -\Lambda \subset \Lambda - S$, they are both
Meyer sets. The Meyer concept of quasicrystals thus generalizes
the classical definition of crystals. Perfect crystal is however a
periodic set, i.e.\ $\Lambda + x \subset\Lambda$ for any $x\in L$,
and therefore it is not a suitable model for quasicrystalline
materials, which reveal rotational symmetries incompatible with
periodicity. We shall now describe a large class of Meyer sets
which are not invariant under translation.

Let $L$ be a full rank lattice in $\mathbb{R}^d$ and let
$\mathbb{R}^d$ be written as a direct sum $V_1\oplus V_2$ of two
subspaces. One of the subspaces, say $V_1$, plays the role of the
space onto which the lattice $L$ is projected, we call $V_1$ the
physical space, the other subspace, $V_2$, determines the
direction of the projection map. $V_2$ is called the inner space.
Let us denote by $\pi_1$ the projection map on $V_1$ along $V_2$,
and analogically for $\pi_2$.

We further require that the full rank lattice is in general
position, it means that $\pi_1$ is one-to-one when restricted to
$L$, and that the image of the lattice $L$ under the projection
$\pi_2$ is a set dense in $V_2$. The situation can be diagrammed
as follows:

\begin{center}
\begin{picture}(100,24)
 \put(50,-5){\makebox(0,0){$L$}}
 \put(50,5){\makebox(0,0){$\cup$}}
 \put(52,17){\makebox(0,0){$\mathbb{R}^d$}}
 \put(40,17){\vector(-1,0){24}}
 \put(8,17){\makebox(0,0){$V_1$}}
 \put(59,17){\vector(1,0){24}}
 \put(92,17){\makebox(0,0){$V_2$}}
 \put(33,22){\makebox(0,0){$\pi_1$}}
 \put(72,22){\makebox(0,0){$\pi_2$}}
\end{picture}
\end{center}

For the definition of $C\& P$ sets we need also a bounded set
$\Omega \in V_2$, called {\it acceptance window}, which realizes
the "cut". The $C\& P$ set is then defined as
$$
\Sigma(\Omega): = \left\{ \pi_1(x) \mid x\in L\quad {\rm and}
 \quad \pi_2(x) \in \Omega \right\}\,.
$$
 A cut-and-project set $ \Sigma(\Omega)$ with acceptance window $\Omega$ is
 formed by lattice points projected on $V_1$, but only those whose projection
on $V_2$ belongs to $\Omega$, i.e. by projections of lattice
points found in the cartesian product $V_1 \times \Omega$.

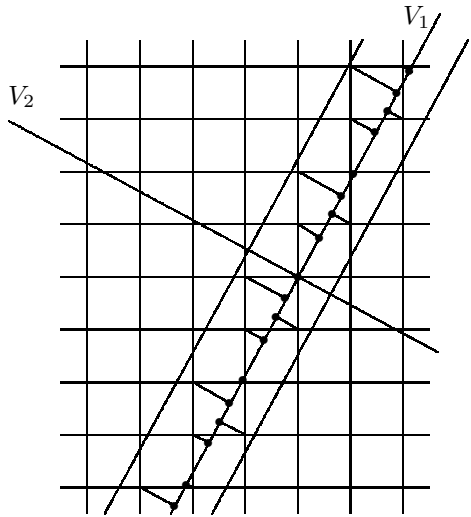
\begin{figure}[hbt]

 \begin{center}
 \setlength{\unitlength}{0.7mm}
 \begin{picture}(105, 95)(-5, 5)
 \put(-5, 83){$V_2$}\put(70,98){$V_1$} \put(5,10){\line(1,0){70}}
 \put(5,20){\line(1,0){70}}\put(5,30){\line(1,0){70}}
 \put(5,40){\line(1,0){70}}\put(5,50){\line(1,0){70}}
 \put(5,60){\line(1,0){70}}\put(5,70){\line(1,0){70}}
 \put(5,80){\line(1,0){70}}\put(5,90){\line(1,0){70}}
 \put(10,5){\line(0,1){90}}
 \put(20,5){\line(0,1){90}}\put(30,5){\line(0,1){90}}
 \put(40,5){\line(0,1){90}}\put(50,5){\line(0,1){90}}
 \put(60,5){\line(0,1){90}}\put(70,5){\line(0,1){90}}
 \qbezier(13.2,5)(40,54)(62.1,95)
 \qbezier(25.75,5)(50,50)(76.9,100)
 \qbezier(-5,79.65)(50,50)(76.5,35.69)
 \qbezier(33.6,5)(60,54)(82.2,95)
 \qbezier(50,40)(47.9,41.25)(45.8,42.5)
 \put(45.8,42.5){\circle*{1.5}} \put(50,50){\circle*{1.5}}
 \qbezier(50,60)(52,58.75)(54,57.5) \put(54,57.5){\circle*{1.5}}
 \qbezier(50,70)(54.25,67.65)(58.5,65.3)
 \put(58.2,65.5){\circle*{1.5}}
 \qbezier(60,60)(58.25,61)(56.5,62)\put(56.5,62){\circle*{1.5}}
 \qbezier(60,70)(60.55,69.75)(60.9,69.5)\put(60.5,69.6){\circle*{1.5}}
 \qbezier(60,80)(62.3,78.75)(64.6,77.5)\put(64.6,77.5){\circle*{1.5}}
 \qbezier(60,90)(64.5,87.5)(69,85)\put(68.7,85){\circle*{1.5}}
 \qbezier(70,80)(68.5,80.75)(67,81.5)\put(67,81.5){\circle*{1.5}}
 \qbezier(70,90)(70.75,89.5)(71.5,89)\put(71.15,89.25){\circle*{1.5}}
 \qbezier(40,20)(37.5,21.25)(35,22.5)\put(35.1,22.5){\circle*{1.5}}
 \qbezier(40,30)(39.5,30.25)(39,30.5)\put(39.5,30.5){\circle*{1.5}}
 \qbezier(40,40)(41.75,39)(43.5,38)\put(43.5,38){\circle*{1.5}}
 \qbezier(40,50)(43.75,48)(47.5,46)\put(47.5,46){\circle*{1.5}}
 \qbezier(30,30)(33.5,28)(37,26)\put(37,26){\circle*{1.5}}
 \qbezier(30,20)(31.5,19.25)(33,18.5)\put(33,18.5){\circle*{1.5}}
 \qbezier(30,10)(29.25,10.25)(28.5,10.5)\put(28.8,10.5){\circle*{1.5}}
 \qbezier(20,10)(23.25,8.25)(26.5,6.5)\put(26.5,6.5){\circle*{1.5}}
 \end{picture}
\caption{Construction of a one-dimensional cut-and-project set.}
\label{fig2}
 \end{center}
 \end{figure}

Figure~\ref{fig2} shows the construction of a $C\& P$ set with
one-dimensional physical and one-dimensional inner space. The
acceptance window here is an interval in $V_2$. And the cylinder
$V_1 \times \Omega$ is a strip in the plane.

Let us list several important properties of $C\& P$
sets:
\begin{itemize}
  \item $\Sigma(\Omega) + t \not\subset  \Sigma(\Omega)$  for any
   $t \in V_1$, i.e. $\Sigma(\Omega)$ is not an ideal crystal.

   \item If the interior $\Omega^\circ$ of the acceptance window is not empty,
   then $\Sigma(\Omega)$ is a Meyer set.

   \item If $\widetilde{\Sigma}$ is a Meyer set, then there exists a $C\&P$ set
   $\Sigma(\Omega)$ with $\Omega^\circ \neq \emptyset$ and
   a finite set $F$ such that $\widetilde{\Sigma}\subset \Sigma(\Omega)
   +F$.
\end{itemize}

First two properties can be derived directly from the definition
of a $C\& P$ set; the third one has been shown in~\cite{Lagarias}.

The aim of physicists is to find a mechanism which forces atoms in
the quasicrystalline matter to occupy given positions. All
physical approaches for describing crystals are based on
 minimum energy argument. If one wants at least to have a chance to find
 a physical explanation why
a given  Delone set is a suitable model for a quasicrystal then
the number of various neighborhoods of atoms (of points) in the
Delone set must be finite.  This requirement is formalized in the
notion of finite local complexity: We say that a Delone set
$\Sigma$ is of {\it finite local complexity} if for any  fixed
radius $r$ all balls of this radius $r$ contain only finitely many
different configurations of points of $\Sigma$ up to translation.

It follows from their definition, that Meyer sets have finite
local complexity. Therefore the condition $\Omega^\circ \neq
\emptyset$ ensures that a $C\& P$ set $\Sigma(\Omega)$ has finite
local complexity.

Another physically reasonable requirement on the model of
quasicrystal is that every configuration of points are found in
the modeling set infinitely many times. This property may be for
example ensured by the requirement that the boundary of the
acceptance window has an empty intersection with the image of the
lattice by the projection, i.e.\ $\partial \Omega \cap \pi_2(L)=
\emptyset$.

\section{Cut-and-project Sets with Rotational Symmetry}\label{sec:rot}

Recalling the motivation for introducing the notion of
quasicrystals, one should ask about conditions ensuring existence
of a crystallographically forbidden symmetry. For this, conditions
on the acceptance window alone are not sufficient. The
construction of a two-dimensional $C\&P$ set with 5-fold symmetry
has been described by Moody and Patera in~\cite{MoodyPatera}.
In~\cite{BCG} one can find a more general construction of $C\&P$
sets with rotational symmetry of order $2n+1$, for $n\in
\mathbb{N}, n\geq 2$.

Consider the lattice $L \subset \mathbb{R}^4$ to be generated by
unit vectors $\alpha_1, \ldots, \alpha_4$ whose mutual position is
given by the following diagram.
\begin{center}
\begin{picture}(110,20)
\put(5,5){\makebox(0,0){$A_4\equiv$}} \put(25,5){\circle{10}}
\put(30,5){\line(1,0){16}} \put(51,5){\circle{10}}
\put(56,5){\line(1,0){16}} \put(77,5){\circle{10}}
\put(82,5){\line(1,0){16}} \put(103,5){\circle{10}}
 \put(27,15){\makebox(0,0){$\alpha_1$}}
 \put(53,15){\makebox(0,0){$\alpha_2$}}
 \put(78,15){\makebox(0,0){$\alpha_3$}}
 \put(105,15){\makebox(0,0){$\alpha_4$}}
\end{picture}
\end{center}
In the diagram the vectors connected by an edge make an angle
$\pi/3$, otherwise are orthogonal. Such vectors are root vectors
of the group which is in physics denoted by $A_4$. It can be
verified that the lattice generated by $\alpha_1, \ldots,
\alpha_4$ is invariant under 5-fold rotational symmetry. Let us
mention that dimension 4 is the smallest which allows a lattice
with such a rotational symmetry.

The physical space $V_1$ and the inner space $V_2$ in our example
have both dimension 2, thus they are spanned by two vectors, say
$v,u$ and $v^\star, u^\star$ respectively. We can choose them as
unit vectors, such that $u$ and $v$ form an angle $\frac{4\pi}{5}$
and $u^\star$ a $v^\star$ form an angle $\frac{2\pi}{5}$. The
following scheme shows the definition of the projection $\pi_1$,
which is uniquely given if specified on the four basis vectors
$\alpha_1, \ldots, \alpha_4$.

\begin{center}
\begin{picture}(253,58)
\put(5,28){\makebox(0,0){$A_4\equiv$}} \put(25,15){\circle{10}}
\put(25,41){\circle{10}} \put(25,20){\line(0,1){16}}
\put(29,38){\line(1,-1){19}} \put(51,15){\circle{10}}
\put(51,41){\circle{10}} \put(51,20){\line(0,1){16}}
 \put(66,28){\vector(1,0){20}}
 \put(77,35){\makebox(0,0){$\pi_1$}}
\put(101,15){\circle{10}} \put(101,41){\circle{10}}
\put(101,20){\line(0,1){16}} \put(105,38){\line(1,-1){19}}
\put(127,15){\circle{10}} \put(127,41){\circle{10}}
\put(127,20){\line(0,1){16}}
 \put(27,4){\makebox(0,0){$\alpha_1$}}
 \put(27,51){\makebox(0,0){$\alpha_2$}}
 \put(53,4){\makebox(0,0){$\alpha_3$}}
 \put(53,51){\makebox(0,0){$\alpha_4$}}
 \put(101,4){\makebox(0,0){$u$}}
 \put(101,51){\makebox(0,0){$\tau v$}}
 \put(127,4){\makebox(0,0){$\tau u$}}
 \put(127,51){\makebox(0,0){$v$}}
 \put(220,20){\makebox(0,0){$\hbox{{where }} \ \ \tau = \frac{1+\sqrt{5}}{2}$\,.}}
\end{picture}
\end{center}

The irrational number $\tau$, usually called the golden ratio, is
the greater root of the quadratic equation $x^2=x+1$. Recall that
regular pentagon of side-length 1 has the diagonal of length
$\tau$, whence the correspondence of the golden ratio with the
construction of a point set having 5-fold rotational symmetry.

The projection $\pi_2$ is defined analogically, substituting
vectors $u,v$ in the diagram by $v^\star, u^\star$, and the scalar
factor $\tau$ by $\tau'=\frac12(1-\sqrt5)$, which is the other
root of the quadratic equation $x^2=x+1$.

With this choice of projections $\pi_1$ and $\pi_2$, a point of a
lattice given by four integer coordinates $(a,b,c,d)$ is projected
as
$$
\begin{array}{rcl}
\pi_1(a,b,c,d) &=& (a+\tau b)v + (c+\tau d)u\,,\\
\pi_2(a,b,c,d) &=& (a+\tau' b)v^* + (c+\tau d')u^*\,,
\end{array}
$$
and the $C\& P $ set has the form
$$
\Sigma(\Omega) =\bigl\{ (a+\tau b)v + (c+\tau d)u\ \bigm|\ a,b,c,d
\in \mathbb{Z}, \
   (a+\tau' b)v^* + (c+\tau d')u^* \in \Omega\bigr\}\,.
$$

\begin{figure}[!h]
    \begin{center}
    \resizebox{11.5cm}{!}{\includegraphics{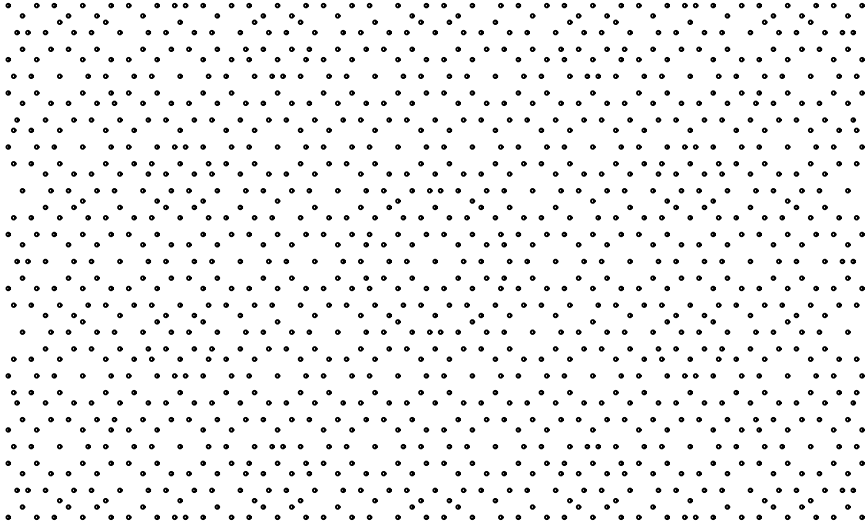}}
    \caption{Two-dimensional cut-and-project set with disc acceptance window.}
    \label{fig3}
    \end{center}
\end{figure}

To complete the definition of the  $C\& P $ set $\Sigma(\Omega)$
we have to provide the acceptance window $\Omega$. Its choice
strongly influences geometrical properties of $\Sigma(\Omega)$.
In~\cite{nejakyMoody} it is shown that with the above
cut-and-project scheme the set $\Sigma(\Omega)$ has 10-fold
rotational symmetry if and only if the 10-fold symmetry is
displayed by the acceptance window $\Omega$. Figure~\ref{fig3}
shows a cut-and-project set $\Sigma(\Omega)$ where $\Omega$ is a
disc centered at the origin.

\medskip

If one studies the inter-atomic interactions, it is impossible to
consider contribution of all atoms in the matter; one must limit
the consideration to `neighbours' of a given atom. Thus it is
necessary to define the notion of neighbourhood in a general point
set, which has not a lattice structure. A natural definition of
neighbours is given in the notion of a Voronoi cell.

Consider a Delone set $\Sigma\subset{\mathbb R}^d$ and choose a
point $x$ in it. The Voronoi cell of $x$ is the set
$$
V(x) = \{y\in{\mathbb R}^d \mid \|x-y\|\leq \|z-y\| \hbox{ for all
 } z\in\Sigma \}\,.
$$

The Voronoi cell of the point $x$ is thus formed by such part of
the space, which is closer to $x$ than to any other point of the
set $\Sigma$. Since $\Sigma$ is a Delone set, the Voronoi cells of
all points are well defined convex polytopes in ${\mathbb R}^d$,
filling this space without thick overlaps. Therefore they form a
perfect tiling of the space.

\begin{figure}[!h]
    \begin{center}
    \resizebox{11.5cm}{!}{\includegraphics{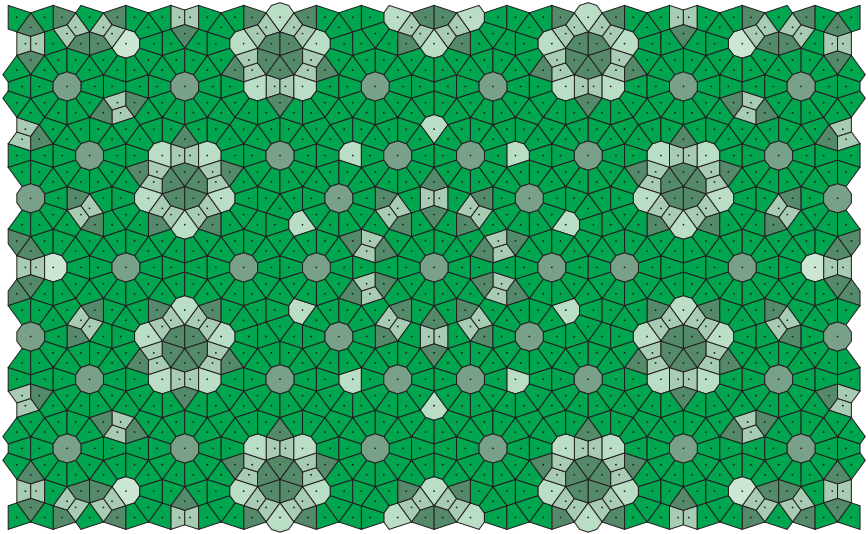}}
    \end{center}
    \caption{Voronoi tiling of the cut-and-project set shown in Figure~\ref{fig3}.}
    \label{fig4}
\end{figure}

The notion of Voronoi cells allows a natural definition of
neighbourhood of points in a Delone set $\Sigma\subset\R^d$. Two
points may be claimed neighbours, if their Voronoi polytopes share
a face of dimension $d-1$. The Voronoi tiling of the
cut-and-project set $\Sigma(\Omega)$ from Figure~\ref{fig3} is
shown on Figure~\ref{fig4}.

In the Voronoi tiling of Figure~\ref{fig4} one finds only 6 basic
types of tiles (Voronoi polygones). They are all found together
with their copies rotated by angles $\frac{\pi}{10}j$,
$j=0,1,\dots,9$.

\begin{figure}[!ht]
    \begin{center}
    \resizebox{11.5cm}{!}{\includegraphics{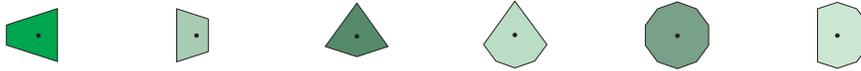}}
    \end{center}
    \caption{The tiles appearing in the Voronoi tiling of Figure~\ref{fig4}.}
    \label{fig5}
\end{figure}

Let us mention that the Voronoi tiling shown at Figure~\ref{fig5}
is aperiodic, since the $C\& P $ set $\Sigma(\Omega)$ is
aperiodic. For certain classes of acceptance windows with 10-fold
rotational symmetry, the collections of appearing Voronoi tiles
are described in the series of articles~\cite{MasakPaterZich1}.

The geometry of the Voronoi tilings generated by cut-and-project
sets is, except several special cases, unknown. The only known
fact is that the number of types of tiles is for every
cut-and-project set finite, which follows from the finite local
complexity of cut-and-project sets. The situation in dimenions
$d\geq2$ is quite complicated. Therefore we focus in the remaining
part of the paper on one-dimensional cut-and-project sets.

\section{One-dimensional $C\&P$ Sets and $C\&P$ Words}\label{sec:def}

Let us describe in detail the construction of a one-dimensional
$C\&P$ set, as illustrated on Figure~\ref{fig2}.
 Consider the lattice $L=\mathbb{Z}^2$ and  two distinct
straight lines $V_1: y=\varepsilon x$ and $V_2:y=\eta x$,
$\varepsilon\neq\eta$. If we choose vectors  \ $
\vec{x}_1=\frac{1}{\varepsilon-\eta}(1,\varepsilon)\ \hbox{ and }\
\vec{x}_2=\frac{1}{\eta-\varepsilon}(1,\eta) $  \ then for any
point of the lattice $\mathbb{Z}^2$ we have
$$
(p,q)=\underbrace{(q-p\eta)\vec{x}_1}_{\pi_1(p,q)} +
\underbrace{(q-p\varepsilon)\vec{x}_2}_{\pi_2(p,q)}\,.
$$
Let us recall that the construction by cut and projection requires
that the projection $\pi_1$ restricted to the lattice $L$ is
one-to-one, and that the set $\pi_2(L)$ is dense in $V_2$.

If $\eta$ and $\varepsilon$ are irrational numbers, then these
conditions are satisfied. The projection of the lattice
$L=\mathbb{Z}^2$ on the straight lines $V_1$ and $V_2$ are written
using additive abelian groups
$$
\begin{array}{rcl}
\mathbb{Z}[\eta]&:=&\{a+b\eta\mid a,b\in\mathbb{Z}\}\,,\\[2mm]
\mathbb{Z}[\varepsilon]&:=&\{a+b\varepsilon\mid
a,b\in\mathbb{Z}\}\,.
\end{array}
$$
These groups are obviously isomorphic; the isomorphism $\star:
\mathbb{Z}[\eta] \to \mathbb{Z}[\varepsilon]$ is given by the
prescription
$$
x=a+b\eta \quad\mapsto\quad x^\star=a+b\varepsilon\,.
$$
The cut-and-project scheme can then be illustrated by the
following diagram.
\begin{center}
\begin{picture}(100,58)
 \put(50,50){\makebox(0,0){$\mathbb{Z}^2$}}
 \put(40,40){\vector(-1,-1){24}}
 \put(8,8){\makebox(0,0){$\mathbb{Z}[\eta]\vec{x}_1$}}
 \put(56,40){\vector(1,-1){24}}
 \put(91,8){\makebox(0,0){$\mathbb{Z}[\varepsilon]\vec{x}_2$}}
 \put(49,14){\makebox(0,0){$\star$}}
 \put(25,8){\vector(1,0){48}}
 \put(18,31){\makebox(0,0){$\pi_1$}}
 \put(78,31){\makebox(0,0){$\pi_2$}}
\end{picture}
\end{center}
Every projected point $\pi_1(p,q)$ lies in the set
$\mathbb{Z}[\eta]\vec{x}_1$. The notation will be simplified by
omitting the constant vector $\vec{x}_1$. In a similar way, we
omit the vector $\vec{x}_2$ in writing the projection
$\pi_2(p,q)$. With this convention, one can define a
one-dimensional cut-and-project set as follows.

\begin{de}
Let $\varepsilon, \eta$ be distinct irrational numbers, and let
$\Omega \subset \mathbb{R}$ be a bounded interval. Then the set
$$
\Sigma_{\varepsilon,\eta}(\Omega):=\{a+b\eta\mid
a,b\in\mathbb{Z},\ a+b\varepsilon\in\Omega\} =\{x\in
\mathbb{Z}[\eta]\mid x^\star\in\Omega\}
$$
is called a one-dimensional cut-and-project set  with parameters
$\varepsilon$, $\eta$ and $\Omega$.
\end{de}

From the properties of general cut-and-project sets, in particular
from their finite local complexity, we derive that the distances
between adjacent points of $\Sigma_{\varepsilon,\eta}(\Omega)$ are
finitely many. The following theorem~\cite{GuMaPe} limits the
number of distances to three.

\begin{thm}\label{mezery}
 For every
$\Sigma_{\varepsilon,\eta}(\Omega)$ there exist positive numbers
$\Delta_1$, $\Delta_2\in\Z[\eta]$ such that the distances between
adjacent points in $\Sigma_{\varepsilon,\eta}(\Omega)$ take values
in $\{\Delta_1,\Delta_2,\Delta_1+\Delta_2\}$. The numbers
$\Delta_1$, $\Delta_2$ depend only on the parameters
$\varepsilon,\eta$ and on the length $|\Omega|$ of the interval
$\Omega$.
\end{thm}

As the theorem states, the distances $\Delta_1,\Delta_2$ and
$\Delta_1+\Delta_2$ depend only on the length of the acceptance
window $\Omega$ and not on its position in $\mathbb{R}$ or on the
fact whether $\Omega$ is open or closed interval. These properties
can, however, influence the repetitivity of the set
$\Sigma_{\varepsilon,\eta}(\Omega)$. More precisely, they can
cause that the largest or the smallest distance appears in
$\Sigma_{\varepsilon,\eta}(\Omega)$ at one place only.

From the proof of Theorem~\ref{mezery} it follows that if $\Omega$
semi-closed, then $\Sigma_{\varepsilon,\eta}(\Omega)$ is
repetitive. Therefore in the sequel we consider without loss of
generality the interval $\Omega$ of the form $\Omega=[c,c+\ell)$.
Nevertheless, let us mention that even if
$\Sigma_{\varepsilon,\eta}([c,c+\ell))$ is repetitive, the
distances between adjacent points may take only two values, both
of them appearing infinitely many times.

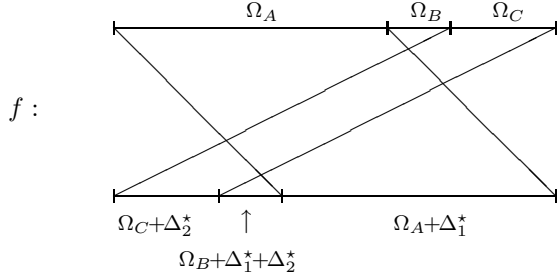
\begin{figure}[ht]
\begin{center}
 {\setlength{\unitlength}{0.28mm}
\begin{picture}(270,130)
\put(30,30){\line(1,0){210}} \put(30,27){\line(0,1){6}}
\put(240,27){\line(0,1){6}} \put(110,27){\line(0,1){6}}
\put(80,27){\line(0,1){6}} \put(30,110){\line(1,0){210}}
\put(30,107){\line(0,1){6}} \put(240,107){\line(0,1){6}}
\put(190,107){\line(0,1){6}} \put(160,107){\line(0,1){6}}
\put(30,110){\line(1,-1){80}} \put(160,110){\line(1,-1){80}}
\put(30,30){\line(2,1){160}} \put(80,30){\line(2,1){160}}
\put(-20,68){$f:$}
 \put(32,14){\scriptsize $\Omega_C\!\! +\!\!\Delta_2^\star$}
\put(61,-3){\scriptsize $\Omega_B\!\!
+\!\!\Delta_1^\star\!\!+\!\!\Delta_2^\star$}
\put(90,15){$\uparrow$}
 \put(163,14){\scriptsize
$\Omega_A\!\!+\!\!\Delta_1^\star$} \put(93,115){\scriptsize
$\Omega_A$} \put(171,115){\scriptsize $\Omega_B$}
\put(210,115){\scriptsize $\Omega_C$}
\end{picture}}
\end{center}
\caption{The diagram illustrating the prescription~\eqref{eq:f}.
The function $f$ is a three-interval exchange transformation.}
\label{fig6}
\end{figure}

The knowledge of values $\Delta_1,\Delta_2$ allows one to easily
determine the neighbours of arbitrary point of the set
$\Sigma_{\varepsilon,\eta}([c,c+\ell))$ and by that, generate
progressively the entire set. Denote $(x_n)_{n\in \mathbb{Z}}$ an
increasing sequence such that
 $\{x_n\mid n\in \mathbb{Z}\} =
\Sigma_{\varepsilon,\eta}([c,c+\ell))$. Then for the images of
points of the set $\Sigma_{\varepsilon,\eta}([c,c+\ell))$ under
the mapping $\star$ one has
 \begin{equation}\label{eq:f}
x_{n+1}^\star=\left\{\begin{array}{ll}
x_n^\star+\Delta^\star_1 &\hbox{ if } \ x_n^\star \in[c,c+\ell-\Delta_1^\star)=:\Omega_A\,,\\
x_n^\star+\Delta^\star_1+\Delta^\star_2 \quad&\hbox{ if }
 \ x_n^\star \in[c+\ell-\Delta_1^\star,c-\Delta_2^\star)=:\Omega_B\,,\\
x_n^\star+\Delta^\star_2&\hbox{ if } \
x_n^\star\in[c-\Delta_2^\star,c+\ell)=:\Omega_C \,.
\end{array}\right.
\end{equation}
The mapping, which to $x^\star_n$ associates the element
$x_{n+1}^\star$ is a piecewise linear bijection $f: [c,c+\ell)
\mapsto [c,c+\ell)$. Its action is illustrated on
Figure~\ref{fig6}. Such mapping is in the theory of dynamical
systems called 3-interval exchange transformation. In case that
$\Delta_1^\star-\Delta_2^\star = |\Omega| = \ell$, the interval
denoted by $\Omega_B$ is empty. This is the situation when the
distances between adjacent points in
$\Sigma_{\varepsilon,\eta}([c,c+\ell))$ take only two values. The
mapping $f$ is then an exchange of two intervals.

In order to record some finite segment of  the set
$\Sigma_{\varepsilon,\eta}([c,c+\ell))$, one can write down to a
list the individual elements of this set. However, this is not the
most efficient way, since every point
$x\in\Sigma_{\varepsilon,\eta}([c,c+\ell))$ is of the form $x  = a
+b\eta \in\Z[\eta]$ and with growing size of the considered
segment, the integer coordinates $a$, $b$, of the points needed
for recording $x= a +b\eta$ grow considerably. Much simpler is to
find a point $x_0\in\Sigma_{\varepsilon,\eta}([c,c+\ell))$ in the
segment and record the sequence of distances between consecutive
points on the left and on the right of $x_0$.

With this in mind, the entire set
$\Sigma_{\varepsilon,\eta}([c,c+\ell))$ can be coded by a
bidirectional infinite word $ (u_n)_{n\in \mathbb{Z}}$ in the
alphabet $\{A,B,C\}$ given as
$$
u_n=\left\{\begin{array}{cl}
A&\hbox{ if }\ x_{n+1}-x_n=\Delta_1\,, \\
B&\hbox{ if }\ x_{n+1}-x_n=\Delta_1+\Delta_2\,,\\
C&\hbox{ if }\ x_{n+1}-x_n=\Delta_2\,.
\end{array}\right.
$$
Such an infinite word is denoted by
$u_{\varepsilon,\eta}(\Omega)$.

\begin{ex}[Mechanical words]\label{ex:mech}
Let us choose irrational $\varepsilon \in (-1,0)$ and irrational
$\eta >0$. We shall consider one-dimensional $C\& P$ set with
acceptance window $\Omega=(\beta-1,\beta]$, for some
$\beta\in\mathbb{R}$. For simplicity of notation we shall put
$\alpha = -\varepsilon \in (0,1)$. From the definition of $C\& P$
sets it follows that
$$
a+b\eta\in\Sigma_{\varepsilon,\eta}(\Omega) \quad
\Leftrightarrow\quad a+b\varepsilon\in\Omega \quad
\Leftrightarrow\quad \beta-1<a-b\alpha\leq\beta \quad
\Leftrightarrow\quad a=\lfloor b\alpha+\beta\rfloor
$$
and therefore the $C\&P$ set is of the form
$$
\Sigma_{-\alpha,\eta}(\beta-1,\beta] = \{\lfloor
b\alpha+\beta\rfloor + b\eta \mid b\in\Z\}\,.
$$
Since $\alpha,\eta>0$, the sequence $x_n:=\lfloor
n\alpha+\beta\rfloor + n\eta$ is strictly increasing and thus the
distances between adjacent points of the $C\&P$ set
$\Sigma_{-\alpha,\eta}(\beta-1,\beta]$ are of the form
\begin{equation}\label{eq:mezery}
x_{n+1}-x_n = \eta + \lfloor (n+1)\alpha+\beta\rfloor - \lfloor
n\alpha+\beta\rfloor =\left\{
\begin{array}{c}
\eta+1 = \Delta_1\,,\\
\eta = \Delta_2\,.
\end{array}\right.
\end{equation}
This $C\&P$ set can therefore be coded by an infinite word
$(u_n)_{n\in \mathbb{Z}}$ in a binary alphabet. Usually one
chooses the alphabet $\{0,1\}$, so that the $n$-th letter of the
infinite word can be expressed as
\begin{equation}\label{eq:mech}
u_n = \lfloor (n+1)\alpha+\beta\rfloor - \lfloor
n\alpha+\beta\rfloor\,.
\end{equation}
Such infinite words $(u_n)_{n\in \mathbb{Z}}$ were introduced
already in~\cite{morse1} and since then, extensively studied,
under the name mechanical words. The parameter $\alpha \in(0,1)$
is called the slope and the parameter $\beta$ the intercept of the
mechanical word.
\end{ex}

\section{Equivalence of One-dimensional $C\&P$ Sets}\label{sec:eqv}

In the previous section we have defined for a triplet of
parameters $\varepsilon$, $\eta$ and $\Omega$ the set
$\Sigma_{\varepsilon,\eta}(\Omega)$ and we have associated to it
the infinite word $u_{\varepsilon,\eta}(\Omega)$. Natural question
is how these objects differ for different triplets of parameters.
Recall that our construction is based on the projection of the
lattice $\mathbb{Z}^2$. The group of linear transformations of
this lattice onto itself is known to have three generators:
$$
G=\bigl\{{\mathbb A}\in M_2(\mathbb{Z}) \,\bigm|\, \det{\mathbb
A}=\pm1\bigr\}= \Bigl\langle \binom{1\ 1}{0\ 1}, \binom{1\ \ 0}{0\
-\!1}, \binom{0\ 1}{1\ 0}\Bigr\rangle
$$
Directly from the definition of $C\&P$ sets it follows that the
action of these three generators on the lattice provides the
identities
 \begin{eqnarray*}
\Sigma_{\varepsilon,\eta}(\Omega)
    &=& \Sigma_{1+\varepsilon,\ 1+\eta}(\Omega)\,, \label{e:trans1}
    \\[1mm]
\Sigma_{\varepsilon,\eta}(\Omega)
    &=& \Sigma_{-\varepsilon,-\eta}(-\Omega)\,, \label{e:trans2}
    \\[1mm]
\Sigma_{\varepsilon,\eta}(\Omega)
    &=&\eta\ \Sigma_{\tfrac1\varepsilon,\tfrac1\eta}
    (\tfrac{1}{\varepsilon}\Omega)\,. 
\end{eqnarray*}

Another identity for $C\& P$ sets is obtained using invariance of
the lattice $\mathbb{Z}^2$  under translations.
$$
a+b\eta+\Sigma_{\varepsilon,\eta}(\Omega) =
\Sigma_{\varepsilon,\eta}(\Omega+a+b\varepsilon)\,,\qquad \hbox{
for any }\ a,b\in\mathbb{Z}\,.
$$

The mentioned transformations were used in~\cite{GuMaPe} for the
proof of the following theorem.

\begin{thm}\label{thmx}
For every irrational numbers $\varepsilon,\eta$,
$\varepsilon\neq\eta$ and every bounded interval $\Omega$, there
exist $\tilde{\varepsilon}$, $\tilde{\eta}$ and an interval
$\tilde{\Omega}$, satisfying
$$
(P) \qquad \tilde{\varepsilon}\in(-1,0), \ \ \tilde{\eta}>0,  \ \
\max(1+\tilde{\varepsilon},-\tilde{\varepsilon}) <
|\tilde{\Omega}| \leq 1
$$
such that
$$
\Sigma_{\varepsilon,\eta}(\Omega) =
s\Sigma_{\tilde{\varepsilon},\tilde{\eta}}(\tilde{\Omega})\,,\qquad
\hbox{ for some }\ s\in\R\,. 
$$
\end{thm}

Multiplying the set
$\Sigma_{\tilde{\varepsilon},\tilde{\eta}}(\tilde{\Omega})$ by a
scalar $s$ can be understood as a choice of a different scale.
From the physical point of view the sets are therefore de facto
the same. In geometry such sets are said to be similar. In the
study of one-dimensional $C\&P$ sets, one can therefore limit the
considerations only to parameters satisfying the condition $(P)$.
One may ask whether the family of parameters can be restricted
even more. More precisely, one asks whether for different triples
of parameters satisfying $(P)$ the corresponding $C\&P$ sets are
essentially different. The answer to such question is almost
always affirmative, except certain, in some sense awkward, cases.
A detailed analysis can be found in~\cite{GuMaPe}.

Theorem~\ref{thmx} concerned geometrical similarity of $C\&P$
sets. If interested only in the corresponding infinite words
$u_{\varepsilon,\eta}(\Omega)$, we can restrict the consideration
even more. This is a consequence of the following two assertions.

\begin{claim}
If the parameters $\varepsilon, \eta_1, \Omega$ and $\varepsilon,
\eta_2, \Omega$ satisfy $(P)$  then the infinite word
$u_{\varepsilon,\eta_1}(\Omega)$ coincides with
$u_{\varepsilon,\eta_2}(\Omega)$.
\end{claim}

Consequently, we can choose the slope of the straight line $V_2$
in the cut-and-project scheme to be $\eta =
-\frac{1}{\varepsilon}$, where $\varepsilon$ is the slope of the
straight line $V_1$. The straight lines $V_1$ and $V_2$ can
therefore be chosen without loss of generality mutually
orthogonal.

\begin{claim}
If the parameters $\varepsilon, \eta, \Omega$ satisfy $(P)$ then
the infinite word $u_{\varepsilon,\eta}(\Omega)$ coincides with
$u_{-1-\varepsilon,\eta}(-\Omega)$ up to permutation of assignment
of letters.
\end{claim}

This statement implies that for the study of combinatorial
properties of infinite words associated with $C\& P$ sets, one can
limit the choice of the parameter $\varepsilon$ to the range
$(-\frac{1}{2}, 0)$.

\section{Factor and Palindromic Complexity of $C\&P$
Words}\label{sec:compl}

For the description of combinatorial properties of infinite words
associated to one-dimensional $C\& P$ sets one uses the
terminology and methods of language theory. Consider a finite
alphabet $\A$ and  a bidirectional infinite word
$u=(u_n)_{n\in\Z}$,
$$
u=\cdots u_{-2}u_{-1}u_0u_1u_2\cdots\,.
$$
The set of factors of $u$ of the length $n$ is denoted
$$
{\mathcal L}_n=\{u_iu_{i+1}\cdots u _{i+n-1}\mid i\in\Z\}\,.
$$
The set of all factors of the word $u$ is the {\em language} of
$u$,
$$
{\mathcal L} = \bigcup_{n\in\N} {\mathcal L}_n\,.
$$
If any factor occurs in $u$ infinitely many times, then the
infinite word $u$ is called  {\it recurrent}. If moreover for
every factor the gaps between its individual occurrences in $u$
are bounded, then $u$ is called {\it uniformly recurrent}.

The factor  complexity of the infinite word $u$ is a mapping
${\mathcal C}:\N\to\N$ such that
$$
{\mathcal C}(n) = \#\{u_iu_{i+1}\cdots u _{i+n-1}\mid i\in\Z\} =
\#{\mathcal L}_n\,.
$$
The complexity of an infinite word  is a measure of ordering in
it: for periodic words it is constant, for random words it is
equal to $(\#\A)^n$.

Since every factor $u_iu_{i+1}\cdots u _{i+n-1}$ of the infinite
word $u=(u_n)_{n\in\Z}$ has at least one extension
$u_iu_{i+1}\cdots u _{i+n}$, it is clear that ${\mathcal C}(n)$ is
a non-decreasing function. If ${\mathcal C}(n) = {\mathcal
C}(n+1)$ for some $n$, then every factor of the length $n$ has a
unique extension and therefore the infinite word $u$ is periodic.
The complexity of aperiodic words is necessarily strictly
increasing function, which implies ${\mathcal C}(n) \geq n+1$ for
all $n$. It is known that mechanical words (defined
by~\eqref{eq:mech}) with irrational slope are aperiodic words with
minimal complexity, i.e.\ ${\mathcal C}(n)= n+1$. Such words are
called bidirectional {\it sturmian} words. A survey of which
functions can express the complexity of some infinite word can be
found in~\cite{Ferenczi}.

The following theorem has been proven in~\cite{GuMaPe} for the
infinite words obtained by the cut-and-project construction.

\begin{thm}
Let $u_{\varepsilon,\eta}(\Omega)$ be a $C\&P$ infinite word  with
$\Omega=[c,c+\ell)$.
 \begin{itemize}
 \item If $\ell\notin\Z[\varepsilon]$, then for any $n\in\N$ we have ${\mathcal C}(n)=2n+1$.

 \item If $\ell\in\Z[\varepsilon]$, then for any $n\in\N$ we have ${\mathcal C}(n)\leq n +const$.
 \end{itemize}
Moreover, the infinite word $u_{\varepsilon,\eta}(\Omega)$ is
uniformly recurrent.
\end{thm}

From the theorem it follows that the complexity of the infinite
word $u_{\varepsilon,\eta}(\Omega)$ depends only on the length
$\ell=|\Omega|$ of the acceptance window and not on its position.
This is the consequence of the fact that language of
$u_{\varepsilon,\eta}(\Omega)$ depends only on $|\Omega|$.

If the parameters $\varepsilon,\eta$ and $\Omega = [c,c+\ell)$
satisfy the condition (P), then the complexity of the infinite
word $u_{\varepsilon,\eta}(\Omega)$ is minimal (i.e.\ ${\mathcal
C}(n)= n + 1$) if and only if $\ell = 1$. Thus the infinite words
$u_{\varepsilon,\eta}([c,c+1))$ are sturmian words. Nevertheless,
the sturmian structure can be found also in words
$u_{\varepsilon,\eta}([c,c+1\ell))$ with $\ell\in\Z[\varepsilon]$.
For, one can prove the following proposition~\cite{GaMaPe}.

\begin{prop}
If  $\ell\in\Z[\varepsilon]$ then there exists a sturmian word
$v=\cdots v_{-2}v_{-1}v_0v_1v_2\cdots$ in $\{0,1\}^\Z$ and finite
words $W_0,W_1$ over the alphabet $\{A,B,C\}$ such that
$$
u_{\varepsilon,\eta}(\Omega) = \cdots
W_{v_{-2}}W_{v_{-1}}W_{v_0}W_{v_1}W_{v_2}\cdots\,.
$$
\end{prop}

The proposition in fact states that the infinite word
$u_{\varepsilon,\eta}(\Omega)$ can be obtained by concatenation of
words $W_0,W_1$ in the order of 0's and 1's in the sturmian word
$v$. Let us mention that Cassaigne~\cite{cassaigne2} has shown a
similar statement for arbitrary one-directional infinite words
with complexity $n+const$. He calls such words quasisturmian.

A reasonable model of quasicrystalline material cannot distinguish
between the ordering of neighbours on the right and on the left of
a chosen atom.  In terms of the infinite word, which codes the
one-dimensional model of quasicrystal, it means that the language
$ {\mathcal L}$ must contain, together with every factor
$w=w_1w_2\ldots w_n$ also its mirror image
$\overline{w}=w_nw_{n-1}\ldots w_1$. The language of $C\& P$ words
satisfies such requirement.

A factor $w$, which satisfies $w= \overline{w}$, is called a {\it
palindrome}, just as it is in natural languages. The study of
palindromes in infinite words has a great importance for
describing the spectral properties of one-dimensional
Schr\"{o}dinger operator, which is associated to $(u_n)_{n\in
\mathbb{Z}}$ in the following way: To every letter of the alphabet
$a\in {\mathcal A}$ one associates the potential $V(a)$ in such a
way that the mapping $V: {\mathcal A} \mapsto \mathbb{R}$ is
injective. The one-dimensional Schr\"{o}dinger operator $H$ is
then defined as
$$
(H\phi)(n) = \phi(n+1) +\phi(n-1) + V(u_n) \phi(n)\,,\qquad
\phi\in\ell^2(\mathbb{Z})\,.
$$

The spectral properties of $H$ influence the conductivity
properties of the given structure. Roughly speaking, if the
spectrum is absolutely continuous, then the structure behaves like
a conductor, while in the case of pure point spectrum, it behaves
like an insulator. In~\cite{HKS} one shows the connection between
the spectrum of $H$ and the existence of infinitely many
palindromes in the word $(u_n)_{n\in \mathbb{Z}}$.

The function that counts the number of palindromes of a given
length in the language $ {\mathcal L}$ of an infinite word $u$ is
called the palindromic complexity of $u$. Formally, the
palindromic complexity of $u$ is a mapping ${\cal P}:\N \mapsto
\N$ defined by
$$
{\cal P}(n):= \#  \{  w \in {\cal L}_n \mid \overline{w} = w\}\,.
$$

Upper estimates on the number ${\cal P}(n)$ of palindromes of
length $n$ in an infinite word $u$ can be obtained using the
factor complexity ${\mathcal C}(n)$ of $u$. In~\cite{ABCD} the
authors prove a result which puts in relation between the factor
complexity ${\mathcal C}(n)$ and the palindromic complexity ${\cal
P}(n)$. For a non-periodic infinite word $u$ it holds that
\begin{equation}\label{abcd}
{\mathcal P}(n) \leq \frac{16}{n}\  {\mathcal
C}\left(n+\lfloor\frac{n}{4}\rfloor\right)\,.
\end{equation}

Combination of the above estimate with the knowledge of the factor
complexity we obtain for $C\&P$ infinite words that ${\cal P}(n)
\leq 48$.

Infinite words constructed by cut and projection are uniformly
recurrent. For such words, the upper estimate of the palindromic
complexity can be improved, using the observation that uniformly
recurrent words have either ${\mathcal P}(n)=0$ for sufficiently
large $n$, or the language $\mathcal L$ of the infinite word is
invariant under the mirror image, see~\cite{Peter+ja}. If
$\mathcal L$ contains with every factor $w$ its mirror image
$\overline{w}$, then
\begin{equation}\label{lepsi}
{\mathcal P}(n) + {\mathcal P}(n+1) \leq 3 \Delta {\mathcal
C}(n):= 3\Bigl({\mathcal C}(n+1)-{\mathcal C}(n)\Bigr)\ .
\end{equation}
This estimate of the palindromic complexity is better than that
of~\eqref{abcd} in case that the factor complexity ${\mathcal
C}(n)$ is subpolynomial. In particular, for $C\& P$ infinite words
we obtain ${\cal P}(n) \leq 6$. In~\cite{FeZa} one can find the
exact value of the palindromic complexity for infinite words
coding three-interval exchange transformation. Since this is the
case of $C\& P$ words, we have the following theorem.

\begin{thm}
Let $u_{\varepsilon,\eta}(\Omega)$ be a  $C\&P$ infinite word with
$\Omega=[c,c+\ell)$  and let $\varepsilon,\eta, \Omega$ satisfy
the conditions (P). Then
$$
 {\mathcal P}(n)=\left\{
 \begin{array}{cl}
 1 &\ \hbox{for }\ n\  \hbox{even},\\[1mm]
 2 &\ \hbox{for }\ n \ \hbox{odd and }\ \ell = 1,\\[1mm]
 3 &\ \hbox{for }\ n \ \hbox{odd and }\ \ell < 1.
 \end{array}
 \right.
 $$
 \end{thm}

\section{Substitution Invariance of $C\&P$ Words}\label{sec:subst}

To generate the set $\Sigma_{\varepsilon,\eta}(\Omega)$ using the
definition resumes in deciding for every point of the form
$a+b\eta$, whether $a+b\varepsilon$ belongs to the interval
$\Omega$ or not. This is done by verifying certain inequalities
between irrational numbers. If we use a computer working with
finite precision arithmetics, the rounding errors take place and
in fact, the computer generates a periodic set, instead of
aperiodic $\Sigma_{\varepsilon,\eta}(\Omega)$. The following
example gives a hint to much more efficient and in the same time
exact generation of a $C\&P$ set.

Consider the most popular one-dimensional cut-and-project set,
namely the Fibonacci chain. It is a $C\&P$ set with parameters
$\eta = \tau$, $\varepsilon = \tau'$ and $\Omega = [0,1)$. (Recall
that the golden ratio $\tau =\frac{1+\sqrt{5}}{5}$ and $\tau' =
\frac{1-\sqrt{5}}{5}$ are the roots of the equation $x^2=x+1$.)
Since $\tau^2=\tau+1$, the set of all integer combinations of 1
and $\tau$ is the same as the set of all integer combinations of
$\tau^2$ and $\tau$, formally
$$
\tau \Z[\tau] = \Z[\tau]\,.
$$
Moreover, $\Z[\tau]$ is closed under multiplication, i.e.\
$\Z[\tau]$ is a ring. Since $\tau+ \tau'=1$, we have also
$\Z[\tau] = \Z[\tau']$, and the mapping $\star$ which maps
$a+b\tau \mapsto a+b\tau'$ is in fact an automorphism on  the ring
$\Z[\tau]$. Note that  $\Z[\tau]$ is the ring of integers in the
field $\mathbb{Q}[\tau]$ and  $\star$  is the restriction of the
Galois automorphism of this field.

\begin{figure}[ht]
\setlength{\unitlength}{0.31mm}
\begin{center}
\begin{picture}(355,160)
\put(-10,28){\makebox(0,0){$\tau^2\Sigma $\ :}}
\put(4,28){\line(1,0){364}} \put(-10,100){\makebox(0,0){$\Sigma $\
:}} \put(4,100){\line(1,0){364}}
\put(7,25){\line(0,1){80}} \put(23,20){\makebox(0,0){$A$}}
\put(23,107){\makebox(0,0){$A$}} \put(40,25){\line(0,1){80}}
\put(7,28){\circle*{5}} \put(56,20){\makebox(0,0){$A$}}
\put(56,107){\makebox(0,0){$A$}} \put(72,25){\line(0,1){80}}
\put(82,20){\makebox(0,0){$B$}} \put(82,107){\makebox(0,0){$B$}}
\put(92,20){\line(0,1){85}} \put(92,28){\circle*{5}}
\put(92,100){\circle*{5}} \put(108,20){\makebox(0,0){$A$}}
\put(112,107){\makebox(0,0){$A$}} \put(124,25){\line(0,1){80}}
\put(124,100){\circle*{5}} \put(134,20){\makebox(0,0){$B$}}
\put(136,107){\makebox(0,0){$B$}}
\put(144,5){\line(0,1){140}}\put(142,-5){$0$}
\put(144,100){\circle*{5}} \put(160,20){\makebox(0,0){$A$}}
\put(160,107){\makebox(0,0){$A$}} \put(176,25){\line(0,1){80}}
\put(144,144.45){\circle*{5}}
\put(144,28){\circle*{5}} \put(176,100){\circle*{5}}
\put(192,20){\makebox(0,0){$A$}} \put(160,89){\makebox(0,0){\small
$\underbrace{ \hspace*{27pt}}_{\hbox{\footnotesize $\ \
\Delta_1$}}$}}
\put(134,89){\makebox(0,0){\small $\underbrace{}_{\hbox{\footnotesize $\ \Delta_2$}}$}}
\put(189,107){\makebox(0,0){$A$}} \put(208,25){\line(0,1){80}}
\put(208,100){\circle*{5}} \put(218,20){\makebox(0,0){$B$}}
\put(218,107){\makebox(0,0){$B$}} \put(228,20){\line(0,1){85}}
\put(228,100){\circle*{5}} \put(244,20){\makebox(0,0){$A$}}
\put(244,107){\makebox(0,0){$A$}} \put(260,25){\line(0,1){80}}
\put(228,28){\circle*{5}}
\put(276,20){\makebox(0,0){$A$}} \put(276,107){\makebox(0,0){$A$}}
\put(292,25){\line(0,1){80}} \put(302,20){\makebox(0,0){$B$}}
\put(302,107){\makebox(0,0){$B$}} \put(312,25){\line(0,1){80}}
\put(312,28){\circle*{5}} \put(328,20){\makebox(0,0){$A$}}
\put(328,107){\makebox(0,0){$A$}} \put(344,25){\line(0,1){80}}
\put(354,20){\makebox(0,0){$B$}} \put(354,107){\makebox(0,0){$B$}}
\put(364,25){\line(0,1){80}} \put(364,28){\circle*{5}}
\qbezier(144,144.45)(92,100)(7,28)
\qbezier(144,144.45)(124,100)(92,28)
\qbezier(144,144.45)(176,100)(228,28)
\qbezier(144,144.45)(208,100)(312,28)
\qbezier(144,144.45)(228,100)(364,28) \put(186,5){\makebox(0,0)
{\small $\underbrace{\hspace*{72pt}}_{\hbox{$\tau^2\Delta_1$}}$}}
\put(118,5){\makebox(0,0) {\small
$\underbrace{\hspace*{44pt}}_{\hbox{$\tau^2\Delta_2$}}$}}
\end{picture}
\end{center}
\caption{Selfsimilarity and substitution invariance of the
Fibonacci word.} \label{fig7}
\end{figure}
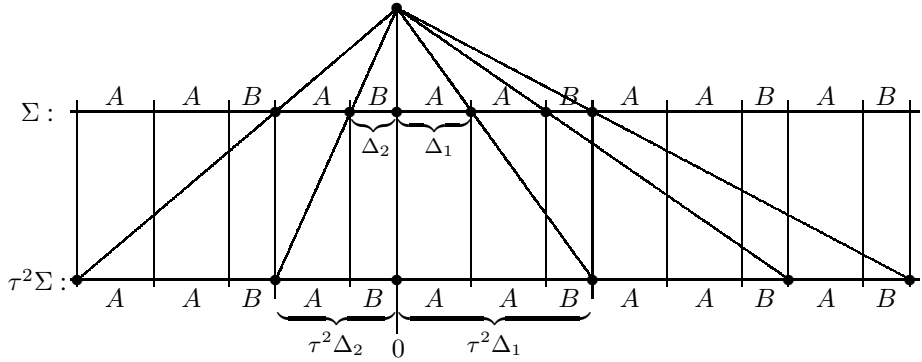

 Using the mentioned properties one can
derive directly from the definition of $C\&P$ sets that
$$
\tau^2\Sigma_{\tau',\tau}(\Omega) =
\Sigma_{\tau',\tau}\Bigl((\tau')^2\Omega\Bigr)\,,
$$
which is valid for every acceptance window $\Omega$. In the case
$\Omega=[0,1)$ we moreover have ${(\tau')^2}\Omega \subset
\Omega$. Therefore
$$
\tau^2\Sigma_{\tau',\tau}(\Omega) \subset
\Sigma_{\tau',\tau}(\Omega)\,,
$$
i.e.\ $\Sigma_{\tau',\tau}(\Omega)$ is selfsimilar with the
scaling factor $\tau^2$, as illustrated on Figure~\ref{fig7}.

Example~\ref{ex:mech}, namely equation~\eqref{eq:mezery} implies
that $\Sigma_{\tau',\tau}(\Omega)$ has two types of distances
between adjacent points, namely $\Delta_1 = \tau^2$ and
$\Delta_2=\tau$. In Figure~\ref{fig7} the distance $\Delta_1$ is
coded by the letter $A$ and the distance $\Delta_2$ by the letter
$B$.

For our considerations it is important that every distance $A$
scaled by the factor $\tau^2$ is filled by two distances $A$
followed by $B$. Similarly, every scaled distance $B$ is filled by
$A$ followed by $B$. This property can be proven from the
definition of $\Sigma_{\tau',\tau}(\Omega)$.

As a consequence, the Fibonacci chain can be generated by taking
an initial segment of the set, scaling it by $\tau^2$ and filling
the gaps by new points in the above described way, symbolically
written as the rule
$$
A \mapsto AAB \qquad  \hbox{and } \qquad B\mapsto AB.
$$
Repeating
this, one obtains step by step the entire $C\&P$ set. Since the
origin $0$ as an element of $\Sigma_{\tau',\tau}(\Omega)$ has its
left neighbour in the distance $\Delta_2$ and the right neighbour
in the distance $\Delta_1$, we can generate the Fibonacci chain
symbolically as
$$
B|A \ \mapsto\  AB|AAB \ \mapsto\ AABAB|AABAABAB \ \mapsto\ \ldots
$$

A natural question arises, whether such efficient and exact
generation is possible also for other one-dimensional
cut-and-project sets, respectively their infinite words. Let us
introduce several notions which allow us to formalize this
question.

A mapping $\varphi$ on the set of finite words over the alphabet
${\mathcal A}$ is a morphism, if the $\varphi$-image of a
concatenation of two words is concatenation of the
$\varphi$-images of the individual words, i.e.\ $\varphi(vw) =
\varphi(v)\varphi(w)$ for every pair of words $v,w$ over the
alphabet $\A$. For the determination of a morphism, it suffices to
specify the $\varphi$-images of the letters of the alphabet. The
action of a morphism can be naturally extended to infinite words,
$$
\varphi(u)=\varphi(\ldots u_{-2}u_{-1}|u_0u_1u_2u_3\ldots):=
\ldots
\varphi(u_{-2})\varphi(u_{-1})|\varphi(u_1)\varphi(u_2)\varphi(u_3)\ldots
$$
An infinite word $u$ invariant under the action of the morphism,
i.e.\ which satisfies $\varphi(u)=u$, is called a fixed point of
$\varphi$. In this terminology, one can say that the Fibonacci
chain (or the infinite word coding it) is a fixed point of the
morphism $\varphi$ over a two-letter alphabet $\{A,B\}$, which is
determined by the images of letters, $\varphi(A) =AAB$,
$\varphi(B) =AB$.

The identity map, which maps every letter of the alphabet on
itself, is also a morphism and arbitrary infinite word is its
fixed point. However, one cannot use the identity map for
generation of infinite words. Therefore we must put additional
requirements on the morphism $\varphi$.

The morphism $\varphi$ over the alphabet ${\mathcal A}$ is called
a substitution, if for every letter $a\in {\mathcal A}$ the length
of the associated word $\varphi(a)$ is at least 1, and if there
exist letters $a_0,b_0\in {\mathcal A}$ such that the words
$\varphi(a_0)$ and $\varphi(b_0)$ have length at least 2, the word
$\varphi(a_0)$ starts with the letter $a_0$, and the word
$\varphi(b_0)$ ends with the letter $b_0$.

A morphism, which is in the same time a substitution, necessarily
has a fixed point $u$, which can be generated by repeated
application of the morphism on the pair of letters $b_0|a_0$.
Formally,
$$
\varphi(u)=u = \lim_{n\to\infty} \varphi^n(b_0)|\varphi^n(a_0)\,.
$$

To every substitution $\varphi$ over the alphabet ${\mathcal{A}} =
\{a_1,a_2, \ldots ,a_k\}$ one associates a  $k\times k$ square
matrix $M$ (the so-called {\it incidence matrix} of the
substitution). The element $M_{ij}$ is given as the number of
letters $a_j$ in the word $\varphi(a_i)$. The incidence matrix of
the substitution generating the Fibonacci word is $M= \left(
\begin{smallmatrix}
 2& 1\\
 1& 1
  \end{smallmatrix}
 \right).$

The incidence matrix of a substitution is by definition a
non-negative matrix for which Perron-Frobenius theorem
holds~\cite{Fiedler}. A substitution $\varphi$ is said {\it
primitive}, if some power of its incidence matrix is positive. In
this case the spectral radius of the matrix is an eigenvalue with
multiplicity 1, the corresponding eigenvector (the so-called
Perron eigenvector of the matrix) being also positive.

Although the incidence matrix $M$ does not allow to reconstruct
the substitution $\varphi$, many properties of the fixed points of
$\varphi$ can be derived from it. Let us mention some of them.

\begin{itemize}
\item If the infinite word $u$ is invariant under a substitution,
then there exists a constant $K$ such that for the complexity
function of the word $u$ we have
$$
{\mathcal C}(n) \leq K n^2 \quad \hbox{for all }\ \ n\in
\mathbb{N}\,.
$$

\item If the infinite word $u$ is invariant under a primitive substitution,
then there exist constants $K_1$ and $K_2$ such that for the
factor complexity and palindromic complexity of the word $u$ we
have
$$
{\mathcal C}(n) \leq K_1 n \quad  \hbox{and} \quad   {\mathcal
P}(n) \leq K_2 \quad  \hbox{for all} \ \ n\in \mathbb{N}\,.
$$

\item An infinite word which is invariant under a primitive
substitution is uniformly recurrent.

\item If the infinite word $u$ is invariant under a primitive substitution
$\varphi$ over an alphabet ${\mathcal A} =
\{a_1,a_2,\ldots,a_k\}$, then every letter $a_i$ has well defined
density in $u$, i.e.\ the limit
$$
\rho(a_i) := \lim_{n\to\infty} \frac{\hbox{number of letters $a_i$
in the word } u_{-n}\ldots u_{-1}|u_0u_1\ldots
 u_{n-1}}{2n+1}
$$
exists. Let $(x_1, x_2,\ldots x_k)$  be the Perron eigenvector of
the matrix $M^\top$ (transpose of the incidence matrix $M$ of the
substitution $\varphi)$. Then the density $\rho(a_i)$ is equal to
$$
\rho(a_i) = \frac{x_i}{x_1+x_2 +\ldots +x_k}\,.
$$
\end{itemize}

The question of describing all $C\&P$ infinite words invariant
under a substitution is still unsolved. A complete solution is
known only for $C\&P$ words over a binary alphabet, which can be,
without loss of generality, represented by mechanical
words~\eqref{eq:mech} with  irrational slope $\alpha\in (0,1)$ and
intercept $\beta \in[0,1)$.

The substitution invariance of mechanical words has first been
solved in~\cite{homog} for the so-called homogeneous mechanical
words, i.e.\ such that $\beta =0$.

\begin{thm}
The homogeneous mechanical word with slope $\alpha\in (0,1)$ is
invariant under a substitution if and only if $\alpha$ is a
quadratic irrational number whose conjugate $\alpha'$ does not
belong to $(0,1)$.
\end{thm}

A quadratic irrational number $\alpha \in (0,1)$ whose conjugate
$\alpha'\notin (0,1)$ is called  {\it Sturm number}. Let us
mention that in the paper~\cite{homog} the Sturm number is defined
using its continued fraction expansion. The simple algebraic
characterization was given in~\cite{allauzen}.


The substitution invariance for general (inhomogeneous) mechanical
words is solved in \cite{sturmian} and \cite{yasutomi}.

\begin{thm}
Let $\alpha$ be an irrational number, $\alpha\in(0,1)$, and let
$\beta\in[0,1)$. The mechanical word with slope $\alpha$ and
intercept $\beta$ is invariant under a substitution if and only if
the following three conditions are satisfied:
\begin{itemize}
\item[{\rm (i)}] $\alpha$ is a Sturm number,

\item[{\rm (ii)}] $\beta\in\mathbb{Q}[\alpha]$,

\item[{\rm (iii)}] $\alpha'\leq \beta' \leq 1-\alpha'$ or $1-\alpha'\leq \beta' \leq \alpha'$,
where $\beta'$ is the image of $\beta$ under the Galois
automorphism of the field $\mathbb{Q}[\alpha]$.
\end{itemize}
\end{thm}

Unlike the case of binary $C\& P$ words, the question of
substitution invariance for ternary $C\& P$ words has been sofar
solved only partially. The following result is the consequence
of~\cite{adam,boshernitzan}.

\begin{thm}
Let $\Omega = [c,d)$ be a bounded interval. If the infinite word
$u_{\varepsilon,\eta}(\Omega)$ is invariant under a primitive
substitution, then $\varepsilon$ is a quadratic irrational number
and the boundary points $c$, $d$ of the interval $\Omega$ belong
to the quadratic field $\mathbb{Q}(\varepsilon)$.
\end{thm}

All $C\& P$ words satisfying the properties of the theorem have a
weaker property than substitution invariance, the so-called
substitutivity~\cite{durand}. It allows one to generate even those
infinite words which are not fixed points of a morphism.

\section{Conclusions}

In the theory of mathematical quasicrystals, best known are the
properties of the one-dimensional models, be it the geometric or
the combinatorial aspects of these structures. However, this
information can be used also in the study of higher dimensional
models, since the one-dimensional ones are embedded in them. In
fact, every straight line containing at least two points of a
higher-dimensional cut-and-project set, contains infinitely many
of them, and they ordering is a one-dimensional cut-and-project
sequence.

Nevertheless, the notions of combinatorics on words, as they were
presented here, are being generalized also to higher dimensional
structures; for example, one speaks about complexity and
substitution invariance of two-dimensional infinite words, even
two-dimensional sturmian words are well defined~\cite{vuillon,
v2}.

Except cut-and-project sets, there are other aperiodic structures
which can serve for quasicrystal models; they are based on
non-standard numeration systems~\cite{burdik}. The set of numbers
with integer $\beta$-expansions share many properties required
from one-dimensional quasicrystal models, in particular, they are
Meyer sets, are self-similar, and the corresponding infinite words
are substitution invariant.

\section*{Acknowledgements}

The authors acknowledge partial support by Czech Science
Foundation GA \v CR 201/05/0169 and by the project LC00602 of the
Czech Ministry of Education.


\end{document}